\documentclass[runningheads]{svmult}
\usepackage{makeidx}   % allows index generation
\usepackage{graphicx}  % standard LaTeX graphics tool
\usepackage{subeqnar}  % subnumbers individual equations
\usepackage{multicol}  % used for the two-column index
\usepackage{cropmark} % cropmarks for pages without
\usepackage{physprbb}  % centered layout of diverse elements, etc.
\def\simge{\mathrel{%
   \rlap{\raise 0.511ex \hbox{$>$}}{\lower 0.511ex \hbox{$\sim$}}}}
\def\simle{\mathrel{
   \rlap{\raise 0.511ex \hbox{$<$}}{\lower 0.511ex \hbox{$\sim$}}}}

\def\be{\begin{equation}}
\def\ee{\end{equation}}
\def\bea{\begin{eqnarray}}
\def\eea{\end{eqnarray}}
\def\ba{\begin{array}} %%%%%%%
\def\ea{\end{array}}

\begin{document}

\title*{\LARGE Constraining the strength 
\protect\newline
of \,Dark Matter Interactions
\protect\newline
from \,Structure Formation}

\toctitle{Constraining Dark Matter Candidates from Structure Formation}
%\protect\newline}
% allows explicit linebreak for the table of content

\titlerunning{Constraining Dark Matter Candidates from
Structure Formation}
% allows abbreviation of title, if the full title is too long
% to fit in the running head
%
\author{C\'eline Bo$\!$ehm\inst{1}
\and Pierre Fayet\inst{2}
\and Richard Scha$\!$effer\,$^{\ \ \ *\!\!\!\!\!\!\!\!\!\!\!\!}$\,\inst{3}}
\authorrunning{C. Bo$\!$ehm et al.}
% if there are more than two authors,
% please abbreviate author list for running head
%
%
\institute{Astrophysics Lab., Keble Road, Oxford, UK
\and 
Laboratoire de Physique Th\'eorique de l'Ecole Normale Sup\'erieure,\\
UMR 8549 CNRS, \,24 rue Lhomond, 75231 Paris Cedex 05, France
\and 
SPhT, CEA Saclay, 91191 Gif-sur-Yvette, France}

\maketitle              % typesets the title of the contribution

\vspace{5mm}

\noindent
*\ \ {Presented by Pierre Fayet}

\begin{abstract}

We discuss the damping of primordial dark matter fluctuations, 
taking into account explicitly the interactions of dark matter
\,-- whatever their intensity --\,
both with itself and with other particle species.
Relying on a general classification of dark matter particle candidates,
our analysis provides, from structure formation,
{\it \,a new set of constraints\,} on the dark matter particle mass and 
interaction rates (in particular with photons and neutrinos).

This determines up to which cross sections the dark matter interactions 
may effectively be disregarded, and when they start playing an essential 
r\^ole, either through collisional damping or 
through an {\it enhancement of the free-streaming scale}.
\,It leads us to extend the notions of Cold, Warm and Hot Dark Matter 
scenarios when dark matter interactions are no longer taken 
to be negligible.
It also suggests the possibility of new scenarios
of {\it \,Collisional Warm Dark Matter},
with moderate damping induced by dark matter interactions.

\end{abstract}

\section{Introduction}

We would like to approach here the question of 
the nature of the non-baryonic dark matter of the Universe   
\,-- assuming it is made of particles --\, through the constraints 
which may be imposed, from structure formation, 
on the strength of dark matter interactions.
Beyond the well-known list of dark matter candidates, 
we would like to know, in a more general way, which kinds of particles 
are admissible and which ones ought to be rejected, taking explicitly 
into account the effects of dark matter interactions not only 
with itself, but also with other particle species such as, 
most notably, photons and neutrinos.
This presentation, which summarizes some of the main results obtained 
in Ref.\,\cite{bfs}, relies on a general and systematic 
discussion of dark matter properties in terms of a few specific parameters
(typically its mass and interaction rates), 
independently of any specific underlying particle theory or model one may 
have in mind.

\vskip .4truecm

\noindent
{\small DARK 2002, 
\ 4$^{\rm th}$ Int. Conf. on Dark Matter in Astro and Particle Physics,
Cape Town, Feb. 2002}\\

\vspace*{-.75truecm}
\hfill {\small LPTENS\,-\,02/34}

\newpage

The interactions of dark matter particles were generally taken as negligible
in the past, and therefore disregarded, when discussing structure formation.
But such interactions should normally exist, and are even essential 
when estimating the relic density of dark matter particles 
having a sizeable mass,
which should annihilate at a sufficient rate
(otherwise their relic abundance would be too
large)\footnote{Unless of course these particles may be diluted 
by some appropriate inflation mechanism.}. 
\,The importance of such annihilations is well-known, for example, 
in the familiar case of
``Weakly-Interacting Massive Particles'', the most favored
being the neutralino, whose stability results from
the $R$-parity of supersymmetric theories.
But even if such interactions are considered as ``weak'', 
is it really legitimate to disregard them when discussing the theory 
of structure formation\,?
The possible r\^ole of dark matter interactions 
is precisely what we would like to study, 
in a general way, irrespectively of 
the specific dark matter candidate one may have in mind.

\section{Collisional damping and free-streaming scales}

\vspace{-.05truecm}

Indeed dark matter properties should be such that the primordial 
fluctuations corresponding to objects of masses larger than about 
$\,10^9\,M_{\hbox{\tiny $\odot$}\,}$ \,(i.e. to sizes 
$\,\simge\,$  \,100\, kiloparsecs)\, should not be erased.
\,The damping of the primordial fluctuations 
may be due to the familiar free-streaming 
of dark matter particles; but also, prior to this free-streaming,
to {\it collisional damping\,} effects 
resulting from dark matter interactions, when these are 
explicitly taken into account.
Up to which point is it legitimate to consider dark matter particles 
as effectively free, and  from which values of the cross-sections 
and interaction rates should dark matter interactions start playing
a significant r\^ole\,?

\subsection{Expressions of the damping scales}

The free-streaming scale associated with dark matter particles 
is simply fixed by the distance 
\,-- as evaluated today --\,
travelled by a freely-propagating dark matter particle, i.e.:
\vspace{-.3truecm}
\be
\label{lfs}
l_{\hbox{\scriptsize \,free-streaming}}\ \ \approx \ \ 
\int_{t_{\rm dec}}^{t_{\rm collapse}}
\ \frac{v_{\hbox{\tiny {dm}}}(t)\,dt}{a(t)}\ \ ,
\ee
in which the integral runs from the time $t_{\hbox{\tiny \,dec}}$
at which dark matter particles start propagating freely, 
until the time of the gravitational collapse of the 
primordial fluctuations, 
$\,a(t)\,$ denoting the scale factor of the Universe at time $t$.

\vskip .2truecm

In case one can consider that dark matter particles, alone,
form a one-com\-po\-nent fluid
(i.e. excluding particles of other species that might be in equilibrium 
with dark matter particles, influencing their damping properties), 
the collisional damping scale of dark matter fluctuations 
may be expressed as follows:
\be
\label{colldamp0}
l_{\hbox{\scriptsize \,coll.-damping}}^{\,2}\ \ \approx\ \ 
\int^{t_{\rm dec}}   \ ...\ 
\ 
\frac{v_{\hbox{\scriptsize \,dm}}^{\,2}(t)\ dt}
{\Gamma_{\hbox{\scriptsize \,dm}}(t)\ \,a^2(t)}\ \ \ .  
\ee
Here $v_{\hbox{\scriptsize \,dm}}^{\,}(t)\,$ 
denotes the average quadratic velocity (at time $\,t$)
\,of a dark matter particle, and $\,\Gamma_{\hbox{\scriptsize \,dm}}(t)\,$ the 
corresponding total interaction rate of this particle
(the  \,...\, stand for additional factors
which usually turn out to be of order unity, as well as an overall 
normalisation factor, which we disregard here for simplicity).
The physical interpretation of this formula will be explained shortly.

\vskip .2truecm

However, in a number of situations it is essential 
to take into account that dark matter particles may remain in equilibrium
with particles belonging to other species (generically denoted by an index 
{\small $\,i\,$}), \,up to a decoupling time 
$\,t_{\hbox{\tiny \,dec}}\hbox{\scriptsize\,(dm-$i$)}.$
~It is then possible to view the dark matter particles as forming, 
together with the particles of these other species, 
a single (multicomponent) fluid.
It is the corresponding damping length of this composite fluid 
that will be relevant for us.

\vskip .2truecm

The collisional damping scale of the fluctuations may in fact be decomposed 
as a quadratic sum of different contributions, associated respectively
with each of the species (including itself)
which have been in equilibrium with dark matter before its decoupling.
Indeed at any given time $\,t$, \,the instantaneous contribution 
to the square of the damping length may be viewed a sum
of the corresponding contributions of the individual components of this fluid,
with weight factors proportional to the energy densities $\,\rho_i\,$
of particles of species $\,i$.
\,Integrating upon time up to the decoupling time of dark matter,
we obtain the expression of the collisional damping scale 
$\,l_{\hbox{\scriptsize \,c.d.}}$, \,which may be written
in a slightly simplified way 
as follows (the  \,...\, standing for additional factors 
which are usually of order unity):
\be
\label{colldamp}
l_{\hbox{\scriptsize \,coll.-damping}}^{\,2}\ \ \approx\ \ 
\int^{t_{\hbox{\tiny \,dec}}\hbox{\tiny\,(dark matter)}}   
\!\!\!\!\!\!\!\!\!\!\!\!\!\!\!\!\!\!\!\!\!\!\!\!\!\!\!\!\!\!\!\!\!\!\!\!
\sum_{\ \ \ \ \ \ \ \ \ \ \ \ \ \ \ \hbox{\scriptsize\,species 
$ i\,= \hbox{\tiny $\left\{
\ba{l} \hbox{\tiny \,dark matter} \vspace{.1truecm}\cr 
\hbox{\tiny other species} \cr \ea
\right.$}
$
}} \!\!\!\!\!\!\!\!\!\!
... \ \ \ \frac{\rho_i}{\rho}\ \ \ \frac{v_i^{\,2}(t)\ dt}{\Gamma_i(t)\ \,a^2(t)}\ \ \ ,
\ee
in which $v_i^{\,}(t)\,$ denotes the average quadratic velocity 
(at time $\,t$)
\,of a particle of species $\,i\,$, and $\,\Gamma_i(t)\,$ the 
corresponding total interaction rate.

\vskip .2truecm

The physical meaning of this formula may be understood easily 
from the fact that for a particle of velocity $\,v_i\,$ 
and total interaction rate $\,\Gamma_i\,$
(i.e. average time between two successive collisions 
$\,\tau_i\,=\,\Gamma_i^{\,-1}\,$), 
the corresponding mean free path between two collisions \,-- 
as it would be measured today --\, is $\,{v_i}/{(\Gamma_i\,a)}$,
\,the corresponding contribution to the mean squared distance 
travelled by the particle being $\,{v_i^{\,2}}/{(\Gamma_i^{\,2}\,a^2)}\,$.
\,The average number of such collisions during a time interval $\,\Delta t\,$
may be expressed as $\,\Delta t/\tau_i\,$ i.e. $\,\Delta t\,\Gamma_i$, 
\,resulting in a contribution 
$\,{v_i^{\,2}\ \Delta t}/{(\Gamma_i\,a^2)}\,$ 
to the mean squared distance travelled by the particle 
of species $i$ considered.
\,This contribution should be weighted 
by a factor proportional to the energy density $\,\rho_i\,$
carried by the particles of species $\,i$, \,divided by the total 
energy density of the fluid.
One should then integrate upon time, up to the decoupling time of dark matter,
which leads to equation (\ref{colldamp}).

\vskip .2truecm

The species with which dark matter may be in equilibrium 
include the dark matter itself. 
The corresponding contribution to the damping scale, called 
the ``\underline{self-damping scale}'', is then given 
by the following expression
\be
\label{lsd}
l_{\hbox{\scriptsize \,self-damping}}^{\,2}\ \ \approx\ \ 
\int^{t_{\rm dec.}\hbox{\tiny\,(dark matter)}}
\ \frac{\rho_{\hbox{\tiny dm}}}{\rho}\ \ 
\frac{v_{\hbox{\tiny dm}}^{\,2}(t)\ dt}{\Gamma_{\hbox{\tiny dm}}(t)\ \,a^2(t)}
\ \ \ ,
\ee
in which $\,\Gamma_{\hbox{\tiny dm}}\,$ denotes the {\it \,total\,} 
interaction rate of a dark matter particle 
\linebreak
(both with other ones
and with other particle species).
~In a similar way, all other species (still denoted by the index $\,i\,$)
\,lead collectively to the following contribution to the collisional 
damping scale, that we shall globally refer to 
as the 
``\underline{induced-damping scale}'', \,given by:
\be
\label{lid}
l_{\hbox{\scriptsize \,induced-damping}}^{\,2}\ \ \approx \!\!\!
\sum_{\hbox{\scriptsize $
\ba{c} 
\hbox{other species $i$} \cr \hbox{coupled to dark matter} \cr 
\ea
$
}}
\int^{t_{\rm dec.}\hbox{\tiny\,(dm-$i$)}} 
\ \frac{\rho_i}{\rho}
\ \ \frac{v_i^{\,2}(t)\ dt}{\Gamma_i(t)\ \,a^2(t)}\ \ \ .
\ee

\vspace{-.25truecm}

\subsection{Constraints on the damping scales}

For a dark matter candidate to be acceptable it is necessary 
to require that the primordial fluctuations in the dark matter density
\,(believed to be at the origin of the formation of the structures)\, 
have not been erased as an effect of collisional damping 
and free-streaming.
We shall then obtain constraints on the possible dark matter 
candidates which may be allowed, 
by demanding that both the collisional damping and the 
free-streaming scales be smaller than a scale 
$\,l_{\hbox{\tiny struct.}}\,$, defined as the scale 
associated with the smallest objects known to be of primordial origin. I.e.:
\vspace{-.2truecm}
\be
\label{cont}
\left\{\ \ \ba{ccc} 
l_{\hbox{\scriptsize \,coll.-damping}}\ \,
&\simle& \ \
l_{\hbox{\scriptsize \,struct.}}\ ,
\\ [.1truecm]
l_{\hbox{\scriptsize \,free-streaming}}\ \,
&\simle& \ \
l_{\hbox{\scriptsize \,struct.}}\ .
\ea \right.
\ee

\vspace{.1truecm}

\noindent
Here we shall consider that this scale $\,l_{\hbox{\scriptsize \,struct.}}\,$
is of the order of 100 kiloparsecs, 
corresponding to masses of the order of $\,10^9\,$ solar masses,
e.g. the mass of a small galaxy.
\,Indeed one can evaluate the collapsing mass associated with a scale 
which was, at collapse time, 100 kpc $a_{\rm c}\,$ to be 
\be
\simeq \ \frac{4\,\pi}{3}\ \,
(100\ \hbox{kpc}\ a_{\rm c})^3\ \rho_c\,\Omega_{\rm m}/a_{\rm c}^{\,3}
\ \simeq\ \,1.16\ \,\Omega_{\rm m}\,h_\circ^{\,2}\ \,10^9\ M_\odot\ \ ,
\ee
(i.e. about $.5 \ \Omega_{\rm m}\ 10^9\ M_\odot\,$ for $\,h_0\,\simeq .65$),
$\,\rho_c$ being the critical density of the Universe.
\,Of course if objects of smaller masses were found to be of primordial origin, 
the constraints discussed here, coming from the inequality
(\ref{cont}), could become significantly more stringent.

\vskip .2truecm
In practice we  shall consider that the constraints (\ref{cont}) 
may be expressed as
\be
\!\ 
\left\{ \ \ \,
\ba{lcl}  \\[-.6truecm]
l_{\hbox{\scriptsize \,coll.-damping}} 
&\ \simle \ &  100\ \hbox{kpc}\ \,, \ \ \hbox{i.e.}\ \left\{\ 
\ba{lcl}
l_{\hbox{\scriptsize \,self-damping}} 
&\ \simle \ &  100\ \hbox{kpc}\ \,, \\ [.1truecm]
l_{\hbox{\scriptsize \,induced-damping}}     
&\ \simle \ &  100\ \hbox{kpc}\ \,,
\ea \right.
\\ [-.1truecm]
l_{\hbox{\scriptsize \,free-streaming}}      
&\ \simle \ &  100\ \hbox{kpc}\ \ . \\ [.05truecm]
\ea
\right.
\ee
We shall therefore evaluate and discuss the different
damping scales, in the various possible conceivable situations.
For this purpose we shall be led to introduce of general classification 
of dark matter candidates, according to their mass and interaction rates, 
as discussed later in section \ref{sec:class}.

\subsection{A first comparison of self-damping and free-streaming scales}

It is already instructive,
before presenting such a classification and making more precise statements,
to compare the self-damping and free-streaming scales. 
~The free-streaming scale (\ref{lfs}), 
$\ l_{\hbox{\scriptsize \,free-streaming}}\,\approx \ 
\int_{t_{\hbox{\tiny \,dec}}}^{t_{\hbox{\tiny \,collapse}}}
\ \frac{v_{\hbox{\tiny{dm}}}(t)\ t}{a(t)}\ \frac{dt}{t}\ ,
$
~obtained by integrating 
$\,v_{\hbox{\tiny dm}}/a\,$ between the decoupling time of dark matter 
and the time of the gravitational collapse, 
may be roughly estimated \cite{bond} as 
\be 
\label{lfs2}
l_{\hbox{\scriptsize \,free-streaming}} \ \ \approx\ \ \pi\ \
\hbox{\small Max.}\ 
\left. \frac{v_{\hbox{\tiny dm}} \, t}{a}\ 
\right|_{\,t_{\rm dec}\hbox{\tiny (dm)}}
^{\,t_{\rm collapse}}
\ \ .
\ee

The accumulated collisional damping scale, on the other hand, 
turns out, in the relevant cases, to be dominated by the contribution 
of the late epochs,
for which the time $\,t\,$ gets close to the decoupling time of dark matter.
The interaction rate of a dark matter particle $\Gamma_{\hbox{\tiny dm}}\,$ 
then gets close to the expansion rate of the Universe $\,H=\dot a/a\,$, 
\,taken at the decoupling time of dark matter.
\,The self-damping scale given by eq.\,(\ref{lsd}) then verifies,
approximately,
\be
\label{lsd2}
\ \ \ \ \ \ \ \ 
l_{\hbox{\scriptsize \,self-damping}} \ \ \approx\ \ 
\pi\ \ \sqrt{\ \frac{\rho_{\hbox{\tiny dm}}}{\rho}}\ \,
\left. \frac{v_{\hbox{\tiny dm}} \, t}{a}\ 
\right|_{\,t_{\hbox{\tiny dec}}\hbox{\tiny (dm)}}
\simle\ \ \pi\ \ 
\left. \frac{v_{\hbox{\tiny dm}} \, t}{a}\ 
\right|_{\,t_{\hbox{\tiny dec}}\hbox{\tiny (dm)}}\ .
\ee
It is, at most, of the same order as the free-streaming scale (\ref{lfs2})
\,(provided of course dark matter particles may actually be 
in the free-streaming regime
after their decoupling, 
i.e. provided they actually decouple before the gravitational collapse).
As a result one generally gets the following inequality between the 
self-damping and free-streaming scales:
\be
\label{ineq}
l_{\hbox{\scriptsize \,self-damping}} \ \ \simle\ \ 
l_{\hbox{\scriptsize \,free-streaming}}\ \ ,
\ee
the two scales being in fact of the same order in a number of situations, 
that we shall discuss later in section \ref{sec:class}.

\vskip .2truecm
There is, however, one possible exception to this statement (\ref{ineq}):
if dark matter were sufficiently strongly interacting so as to remain coupled 
until the time of the gravitational collapse, there would be
{\it \,no free-streaming at all\,} of dark matter particles\,! 
\,In this rather extreme case, for which  dark matter would remain
collisional until (and after) collapse, 
the collisional damping scale is the only one to be considered.

\vskip .2truecm
With the exception of this particular situation, the inequality
(\ref{ineq})
would seem, na\"{\i}vely, to indicate that {\it \,no new constraint\,} 
is to be expected from the consideration 
of the collisional scale associated with {\it \,self-damping\,} 
effects, bounded from above by the free-streaming scale, so that these effects 
of the interactions could seem irrelevant.

\vskip .2truecm
This too na\"{\i}ve conclusion, however, should be corrected, because the
evaluation of the free-streaming scale is itself modified, in the presence 
of interactions. 
As a result even by considering {\it \,free-streaming effects only}, 
one may get, in the case of dark matter particles which 
decouple after becoming non-relativistic
(cf. regions II and III to be discussed later),
{\it \,stronger constraints\,} on the dark matter particle mass
by considering its interactions.
\,And this, {\it \,despite the fact that dark matter particles enter later 
in the free-streaming regime}\,!

\vskip .2truecm

Furthermore, returning to self-damping effects it would be incorrect 
to disregard them
as irrelevant, even if strictly-speaking they do not bring in any additional 
constraint as compared to free-streaming.
Because collisional damping effects are at work 
{\it \,before free-streaming\,},
\,the smaller scale fluctuations are in any case erased
by collisional damping, up to a scale 
$\, l_{\hbox{\scriptsize \,coll.-damping}}$ 
\,(with in a number of cases $\,l_{\hbox{\scriptsize self-damping}}$
as large as
$\,l_{\hbox{\scriptsize \,free-streaming}}$),
\,{\it before\,} free-streaming could have a chance to do it.

\section{Classification of dark matter particle candidates}
\label{sec:class}

To go further and specify the explicit expressions 
of the various damping lengths and subsequent constraints 
of the characteristics 
of dark matter particles, we need to know
{\small \,1)\,} whether dark matter particles are still relativistic 
or not at the time of their decoupling; 
~and {\small \,2)\,} whether their non relativistic 
transition and their decoupling transition \,occur 
while the Universe is still radiation-dominated 
\,(with $\,a(t) \,\simeq\,\sqrt{t/t_{\hbox{\tiny rad}}}\,$),
\,or already matter-dominated 
\,(with \,$a(t) \,\simeq\, (t/t_{\hbox{\tiny mat}})^{2/3}\,)$.

\vskip .2truecm

This leads us to introduce a general
classification of the dark matter particle candidates, 
according to these criteria
\,-- i.e. essentially on their mass $\,m_{\hbox{\tiny{dm}}}$ 
and the strength of their interactions,
which involves their interaction rate $\,\Gamma_{\hbox{\tiny{dm}}}$
evaluated at their decoupling time. 
We therefore consider the following three characteristic times:
\be
\left\{\ \ 
\begin{tabular}{lll}
$\,t_{\rm dec}\,,$ && \ \ the decoupling time of dark matter, \\  [.2truecm]
$\,t_{\rm nr}\,, $ && \ \ at which dark matter particles 
become non-relativistic, 
\\  [.2truecm]
$\,t_{\rm eq}\,, $ && \ \ the ``standard'' time of equality between matter 
and radiation,
\end{tabular}
\right.
\ee
the corresponding scale factors being denoted by
$\,a_{\rm dec}\,,\ a_{\rm nr}\,$ and $\ a _{\rm eq}$, 
\,respectively.
$\ a_{\rm dec}\,$ and $\,a_{\rm nr}\,$ are both unknown 
and depend mostly on the strength of the interactions 
of dark matter particles 
and on their mass, respectively.
$\,a_{\rm eq}$ is in fact defined as a fixed reference time,
given as in standard cosmology by 
$a_{\rm eq}= \rho_{\gamma+\nu}(T_\circ)/\rho_{\rm matter}(T_\circ)$,
\,and roughly equal to $\,10^{-4}\,\Omega_{_{\hbox{\tiny matter}}}^{\ -1}$
\,(for a value of the Hubble expansion parameter 
$\,H_\circ \simeq $ \,65 km/s /Mpc).

\vskip .2truecm

{}From the ordering of these three characteristic times or, equivalently, 
of the corresponding scale factors, 
we can define six general categories of particles, labelled from I to VI.
They may be represented graphically in a plane in which the
horizontal axis corresponds to decreasing values of $\,t_{\rm nr}$
\footnote{On the 
{\it \,horizontal\,} axis, we represent the time $\,t_{\rm nr}\,$ 
of the non-relativistic transition of dark matter by plotting
a quantity proportional to 
$\,1/a_{\rm nr}\,$.
$\ t_{\rm nr}\,$ is defined as the moment for which the dark matter 
temperature drops down to
$\,T_{\hbox{\tiny dm}} (t_{\rm nr}) = m_{\hbox{\tiny dm}}/3\,$.
~The dark matter and photon temperatures may be parametrized as 
$\ T_{\hbox{\tiny{dm}}}(t)\,=
\,T_\circ\,/ \,\kappa_{\hbox{\tiny{dm}}}(t)\ a(t)\,$ and 
$\ T(t)\,=\,T_\circ\,/ \,\kappa(t)\ a(t)$, \,respectively
\,(where $\,\kappa_{\hbox{\tiny{dm}}}$ and $\,\kappa$, 
essentially constant by intervals, account for
the effects of the effective numbers of interacting degrees of freedom, 
and $\,T_\circ\simeq 2.73 \,K\,$).
\ 
We ultimately plot, on this horizontal axis, the product 
$\ m_{\hbox{\tiny{dm}}}\,\kappa_{\hbox{\tiny{dm}}}\,=\,3\ T_\circ/a_{\rm nr}\,
\simeq 7\ \,\hbox{eV} \ 10^{-4}/a_{\rm nr}.$
\,The boundary between regions \,IV-V-VI\, and \,I-II-III\, 
corresponds to
$\,a_{\rm nr}=a_{\rm eq} \simeq 10^{-4}\ 
\Omega_{\hbox{\tiny matter}}^{\ -1} $
and therefore 
$\ m_{\hbox{\tiny{dm}}}\,\kappa_{\hbox{\tiny{dm}}}\,\simeq\ 7 \ \hbox{eV}\
\Omega_{\hbox{\tiny matter}}$, so that regions \,I-II-III\, normally
concern particles heavier than a few eV's.},
\,i.e. increasing values of the dark matter particle mass 
$\,m_{\hbox{\tiny dm}}$;
~and the vertical axis to increasing values of the decoupling time 
$\,t_{\rm dec}$
\footnote{The total interaction rate of a dark matter particle
$\,\Gamma_{\hbox{\tiny{dm}}}$, evaluated at the dark matter decoupling time 
$t_{\rm dec}$,
\,is equal to the corresponding value of the Hubble parameter 
$H =\dot a/a\,$. \,The more strongly a dark matter particle is coupled, 
the later it decouples, and the smallest is its total interaction 
rate at decoupling 
$\,\Gamma_{\hbox{\tiny{dm}}}${\scriptsize (dec)} $\equiv H (t_{\rm dec})$\,.
It is \linebreak
natural to normalize the total interaction rate 
$\Gamma_{\hbox{\tiny{dm}}}$ relatively to the inverse of the volume of a 
comoving cell, i.e. relatively to $\,1/a^3$.
\,Increasing values of $\ \Gamma_{\hbox{\tiny{dm}}}\,a^3\,$ 
(evaluated at $\,t_{\rm dec}$)
\,now correspond to more strongly coupled particles, which decouple later.
More specifically, we plot on the {\it vertical\,} axis 
$\,\Gamma_{\hbox{\tiny{dm}}} \,a^3$, 
evaluated at the decoupling time of dark matter
(or collapse time if dark matter particles were still coupled at that epoch).
}.
\,It can be considered, very roughly, as an 
``interaction strength {\it \,versus\,} mass'' \,representation 
plane: cf. Figure 1  of \,Ref.\,\cite{bfs}.

\vskip .2truecm

The first three regions (I to III) in this plane 
correspond to dark matter particles which
get non-relativistic before the standard time of matter/radiation equality
\,(which is in these cases the actual equality time),
i.e. for which
\vspace{-.15truecm}
\be
\hbox{\large $ a_{\rm nr}\ <\ \,a_{\rm eq}$}\ \,
\hbox{ $\simeq\ 10^{-4}\ \Omega_{\hbox{\tiny matter}}^{\ -1}$} 
\ \ \,.
\ee
\vspace{-.55truecm}

\noindent
This corresponds, typically, to dark matter particle candidates heavier 
than a few eV's. Their classification is given in Table \ref{Tab:reg123},
which may be read from bottom to top,
according to increasing interaction strength (and therefore later decoupling)
of dark matter particles.
We also mention for completeness the three other situations 
corresponding to what we call regions IV, V  and VI,
\,for which $\,a_{\rm eq}<\,a_{\rm nr}$.
~This concerns rather unconventional situations, usually excluded, 
of very light dark matter particles of mass $\,\simle\,$ a few eV's, 
\,that would get non-relativistic 
only after the ``standard'' matter-radiation equality time, 
and lead in general to excessive damping.

\begin{table}
\caption{\ Classification of Dark Matter candidates 
according to the strength of their interactions
(for Dark Matter particles heavier than about a few eV's).}
\label{Tab:reg123}
\begin{center}
\vspace{1mm}
{\normalsize Regions I, II  and III\,:\ \ \ \ \ \ \large  
$ a_{\rm nr}\ <\ \,a_{\rm eq}$} \,
\hbox{$\simeq\ 10^{-4}\ \Omega_{\hbox{\tiny matter}}^{\ -1}$} 
\end{center}
%\vspace{1mm}

\begin{tabular}{|c|}
\hline \\  

``\underline{Very strongly interacting}'' \,dark matter particles \\ 
[.1truecm]

\begin{tabular}{ccc}
\\ [-.1truecm]
{$ $\LARGE  $ \uparrow  $}$\!\! $ &
%%%%%%%%%%%%%%%%%%%%%%%%%%%%%%%%%%%%%%%%%%%%%%%%%%%
\begin{tabular}{c}
\hline \\
Region III\,:\ \ \ \ \ \ \ \  
\normalsize{$\ a_{\rm nr}\,<\ a_{\rm eq}\,<\ a_{\rm dec}\ $} 
\\ [.3truecm] 
\ \ 
Dark Matter particles get non-relativistic before equality, \\ [.1truecm]
but \underline{decouple after  equality}\ \ 
\\ [.3truecm]
``very strongly interacting'' dark matter
\\ [.3truecm]
\hline \\
Region II\,:\ \ \ \ \ \ \ \  
\normalsize{$\ a_{\rm nr}\,<\ a_{\rm dec}\,<\ a_{\rm eq}\ $} 
\\ [.3truecm]
\ Dark Matter particles \underline{first get non-relativistic}, 
then decouple, before equality \ 
\\ [.3truecm]
e.g. \ LSP neutralino,\ ...
\\ [.3truecm]
\hline   \\ 
Region I\,:\ \ \ \ \ \ \ \ 
\normalsize{$\ a_{\rm dec}\,<\ a_{\rm nr}\,<\ a_{\rm eq}\ $}  
\\ [.3truecm]
\ Dark Matter particles \underline{first decouple}, 
then get non-relativistic, 
before equality \ 
\\ [.3truecm]
e.g. gravitino of $\,\sim$ keV,  
\,or massive neutrinos slightly heavier than a few eV's
\\ [.3truecm]
\hline 
\end{tabular} 
%%%%%%%%%%%%%%%%%%%%%%%%%%%%%%%%%
& 
\phantom{$\!$\LARGE  $ \uparrow  $}$\!$
\end{tabular}
\\ [-.1truecm] \\

``No interactions'', i.e. \underline{quasi-free} \,dark matter particles \\
[.3truecm]
\hline
\end{tabular}

\end{table}

\section{\sloppy Comparison of the self-damping and free-stream\-ing scales}

\vskip -.05truecm

We now concentrate on regions I, II and III
\,-- in the order of increasing dark matter interaction strength.

\vskip .2truecm

Within region III one may even consider, at least theoretically,
the possibility that dark matter particles might be {\it so strongly 
interacting\,} that they would remain coupled until (and after) the time 
of the gravitational collapse
(for a scale factor $\,a_{\hbox{\tiny{c}}}\,$ 
typically $\,\sim 1/10\,$).
~We shall refer to this specific subdomain of the parameter space as
region III'.
In this case \,\underline{no free-streaming at all}\, of dark matter particles 
is to be considered, the only possible damping of primordial dark matter 
in this extreme case being collisional damping.

\vskip .2truecm

\begin{table}
\caption{\ Comparison of self-damping and free-streaming lengths.}
\label{Tab:compscales}

\vskip .2truecm
\begin{center}
\begin{tabular}{|l|}
\hline \\ [-.05truecm]
\ \ \ Region III'\,:  \ \ \ \ \ \ \ \ \ \ 
collisional-damping only,  no free-streaming\ !\ \ \ \ \ \ 
\\[.35truecm]
\hline
\\
\ \ \ Region III \ (excluding III')\,:
\\[.2truecm]
 \ \ \ \ \ \ \ \ \ \ 
$l_{\hbox{\scriptsize \,self-damping}} \ \ \approx \ \ 
l_{\hbox{\scriptsize \,free-streaming}}\ \,\approx \ \displaystyle\ \pi\ 
\left. \frac{v_{\hbox{\tiny dm}} \, t}{a}\ 
\right|_{\,t_{\hbox{\tiny dec}}}$\ \,
[
{\scriptsize also\ \, 
$\approx \displaystyle \,\pi
\left. \frac{v_{\hbox{\tiny dm}} \, t}{a}\ 
\right|_{\,t_{\hbox{\tiny eq}}}$}
]\ \ \
\\[.5truecm]
\hline 
\\
\ \ \ Region II\,: \ \ \ \ \ \ \ \ \ \ \ 
$l_{\hbox{\scriptsize \,self-damping}} \ \ \simle \ \
l_{\hbox{\scriptsize \,free-streaming}}\ \,\approx \ \displaystyle\ \pi\ 
\left. \frac{v_{\hbox{\tiny dm}} \, t}{a}\ 
\right|_{\,t_{\hbox{\tiny dec}}}$\ \ \ \
\\[.5truecm]
\hline
\\
\ \ \ Region I\,: \ \ \ \ \ \ \ \ \ \ \ \ \ \
$l_{\hbox{\scriptsize \,self-damping}} \ \ <\ \ 
l_{\hbox{\scriptsize \,free-streaming}} \ \,\approx \ \,
\displaystyle 
\pi\  \left. \frac{c\, t}{a}\ 
\right|_{\,t_{\hbox{\tiny nr}}} $
 \\ [.4truecm]
\ \ \ \ \ \ \ \ \ \ \ \ \ \ \ \ \ \ \ 
\small (in connection with the early decoupling of dark matter,
 \\ 
\ \ \ \ \ \ \ \ \ \ \ \ \ \ \ \ \ \ \ \ \ \ \ \ \ \ \ \ \ \ \ \ \ \ \ \  
\small before it gets non-relativistic)
\\ [.3truecm]
\hline
\end{tabular}
\end{center}
\vspace{-.1truecm}
\end{table}

The comparison of the self-damping and free-streaming 
scales 
(cf. eqs.\,(\ref{lfs2}\,-\ref{ineq}) and section \ref{sec:enhan}) 
is given in Table \ref{Tab:compscales}.
~In regions II and III  (excluding III'), for which dark matter particles 
decouple only 
after becoming non-relativistic (but before the gravitational collapse), 
the self-damping and free-streaming scales may be of the same order. 
In this case
\be
\label{coll}
%\framebox [11.5cm]{\rule[-.6cm]{0cm}{1.4cm} 
\begin{tabular}{c}
{\normalsize \underline{\it collisional damping}} \ \ (not free-streaming) 
\\ [.2truecm]
{\it is actually at the origin of the damping of most of the primordial fluctuations}\,!
\end{tabular}
%}
\ee
Collisional damping then appears as an efficient mechanism at work 
to erase the primordial fluctuations,
even in situations where these fluctuations were previously believed 
to be erased by free-streaming effects\,!

\section{The enhancement of the free-streaming scale, 
as a result of the interactions}
\label{sec:enhan}

Furthermore, the explicit expression of the free-streaming scale 
(\ref{lfs}) is itself modified, when the effects of dark matter 
interactions are taken into account. 
Indeed the velocity of a non-relativistic dark matter particle 
of mass $\,m_{\hbox{\tiny {dm}}}$
decreases more slowly with time if it is interacting 
\,($\,v_{\hbox{\tiny {dm}}} \propto \sqrt {\,T_{\hbox{\tiny {dm}}}(t)} 
\propto 1/\sqrt {\,a(t)}\,$)\, than if it propagates freely 
\,($\,v_{\hbox{\tiny {dm}}} \propto 1/a(t)\,$).

\vskip .17truecm

This leads to an enhancement of the free-streaming scale, for 
particles in regions II and III (as compared to region I). Indeed
from the time $\,t_{\rm dec}\,$ such dark matter particles 
start propagating freely (after getting non-relativistic
at $\,t_{\rm nr}$), they actually do so 
with a higher velocity than if they were propagating freely since the 
time $\,t_{\rm nr}\,$ of their non-relativistic transition.
The corresponding enhancement factor of their velocity at and after 
the decoupling time is roughly
$\,\sqrt{\,a_{\rm dec}/a_{\rm nr}}\,$.
As a result the free-streaming scale of a dark matter particle 
of a given mass 
$m_{\hbox{\tiny {dm}}}\,$ 
turns out to be larger in regions II or III (excluding III') than in region I, 
despite the fact that in these regions the particle remains longer 
in the collisional regime\,!

\vskip .19truecm

In region I the integral
(\ref{lfs}) \,(with $\,v_{\hbox{\tiny {dm}}}(t)\,\approx\,c\,a_{\rm nr}/a(t)\,$ 
once the particle gets non-relativistic)\, yields a free-streaming scale
roughly of the order of 
$\ \pi\ c\,t_{\rm nr}/a_{\rm nr}$ 
$ \simeq 
\pi\,c\,t_{\hbox{\tiny rad}}\, a_{\rm nr}\,$. 
\,It increases with $\,t_{\rm nr}$ roughly as $\,\sqrt {\,t_{\rm nr}}\,$, 
\,i.e. as $\,a_{\rm nr}$, \,or $\,1/m_{\hbox{\tiny dm}}$. 

\vskip .19truecm

In region II, the free-streaming scale is roughly of the order of 
$\pi \,v_{\hbox{\tiny dm}} \, t/a\,$ evaluated at 
$\,t_{\rm dec}\,$,
~with 
$ \ v_{\hbox{\tiny dm}}(t_{\hbox{\tiny dec}}) \,\approx 
c\ \sqrt{a_{\rm nr}/a_{\rm dec}}\,$.
~It is therefore of the order of 
$\,\simeq \pi\ c\,t_{\hbox{\tiny rad}}\, 
\sqrt{\,a_{\rm nr}\,a_{\rm dec}}\,$
\,(which may also be expressed as 
$\,\pi\,c\ (t_{\rm nr} t_{\rm dec}/a_{\rm nr}\,a_{\rm dec})^{1/2}$),
\,i.e. larger than in I, precisely by an enhancement factor 
$\approx \sqrt{\,a_{\rm dec}/a_{\rm nr}}\,$.
\,The free-streaming length increases with $\,t_{\rm dec}$ 
roughly as $\,(t_{\rm dec})^{1/4}$, \,or
$\,\sqrt{\,a_{\rm dec}}\,$.

\vskip .15truecm

In an analogous way in region III, the free-streaming scale is still roughly
$\,\approx \pi \,v_{\hbox{\tiny dm}} \, t/a\,$ 
evaluated at $\,t_{\rm dec}$, as in region II. 
\,But we may now evaluate this quantity, just as well,
at the earlier time $\,t_{\rm eq}$.
\,It may then be expressed as 
\linebreak
$\approx\pi\,c\,t_{\hbox{\tiny rad}}\, \sqrt{\,a_{\rm nr}\,a_{\rm eq}}\,$ 
\,(also  
$\,\approx \,\pi\,c\ (t_{\rm nr} t_{\rm eq}/a_{\rm nr}\,a_{\rm eq})^{1/2}$),
\,and is larger than in region I by a factor 
$\,\approx \sqrt{\,a_{\rm eq}/a_{\rm nr}}\,$
\,(the enhancement factor of the velocity evaluated at $\,t_{\rm eq}\,$,
which behaves roughly like $\,(t_{\rm eq}/t_{\rm nr})^{1/4}$). 
\,This enhancement factor may be significantly larger than unity, 
especially in the case of heavy particles, 
which get non-relativistic very early. Altogether:
\vspace{-.03truecm}
\be
\ba{l}
\hbox{\ \ \ \ In regions II and III \ (excl. III'):}
\\ [.18truecm]
\!\hbox{\it
the \,\underline{free-streaming}\, length is actually \,\underline{enhanced}\, 
as an effect of the interactions\,!
}\\  [.03truecm]
\ea
\ee
\noindent
As a result (and despite the fact that the constraints from self-damping 
may be at most as strong as the ones obtained
from free-streaming only,
cf. eqs.\,\hbox{(\ref{lfs2}-\ref{ineq})} and Table \ref{Tab:compscales}),
{\it \,taking explicitly into account the effects of interactions\,} 
finally turns out to be essential, 
in both regions II and III, to derive the constraints on the characteristics 
of dark matter.

\section{Redefining the Cold, Warm and Hot Dark Matter scenarios, 
in the presence of interactions}

In region I, for which dark matter particles are so weakly 
interacting that they decouple before getting non-relativistic,
the discussion is simple, and the result finally does not depend 
on the effects of the interactions. \,Since 
$\
l_{\hbox{\scriptsize \,self-damping}}\,<\,
l_{\hbox{\scriptsize \,free-streaming}}\,
$,
\,the self-damping constraint may be ignored
in view of the free-streaming one.
Furthermore the expression of the free-streaming length itself
is not significantly affected by the effects of the interactions and the exact
decoupling time of dark matter (which occurs before dark matter 
particles get non-relativistic).
The free-streaming scale of a particle in region I finally reads 
\be
\label{lfs3}
l_{\rm \,free\hbox{-}streaming}\ 
\approx\ \pi\,c\ t_{\rm nr}/a_{\rm nr}\ 
\approx\ \pi\ c\,t_{\hbox{\tiny rad}}\ a_{\rm nr}\ \approx\ 
\pi\ c\,t_{\hbox{\tiny rad}}\ 
3\ T_\circ\,/\,m_{\hbox{\tiny{dm}}}\,\kappa_{\hbox{\tiny{dm}}}\ \ .
\ee
It behaves roughly as the inverse of the mass of the dark matter particle.
To get a first idea on the value of this scale we may use
$\ c\,t_{\hbox{\tiny{rad}}}\,\approx \,c\,t_\circ/\sqrt{\,a_{\rm eq}}
\,\approx 
10^{28}\,\hbox{m}
\,\approx $ $3\ 10^8\ \hbox{kpc}$.
\,With $\,3\,T_\circ \simeq $ $7\ 10^{-4}$ eV\, we then get 
$\ l_{\rm \,f.s.} \approx 100$ kpc
 (10 keV/$m_{\hbox{\tiny{dm}}}\,\kappa_{\hbox{\tiny{dm}}})\ $
as a rough order of magnitude estimate.

\vskip .2truecm

Particles heavier than about $\,\sim$ 10 keV, approximately, would then
lead to acceptable values of the free-streaming scale less than about 100 kpc.
In the lower part of the allowed mass range, this corresponds
to ``warm dark matter'' scenarios, as when interactions were disregarded;
and to ``cold dark matter'', in the case of heavier particles.

\vskip .2truecm

We recover at this stage, in the lower part of region I, 
the usual notions of Cold, Warm and Hot dark matter, 
as they may be classified according to the dark matter particle mass 
or to the corresponding value of the free-streaming scale (\ref{lfs}) 
or (\ref{lfs3}).
``Hot dark matter'' 
(made of, e.g., light massive neutrinos, ...) 
is excluded due 
to excessively large free-streaming scales, 
``warm dark matter''  
corresponding 
to free-streaming scales nearly as large as the allowed value
taken of about \linebreak
$\sim\,$ 100 kpc, and ``cold dark matter'' to significantly
smaller values of this scale.

\vskip .2truecm

We are now able to extend these notions by taking into account, 
systematically, the effects of dark matter interactions, which may both 
1) enhance the value of the free-streaming scale and 2) 
provide physically-interesting collisional damping effects, 
either from self-damping or from induced-damping.
The boundary between region I and region II, 
which obeys the equation $\,a_{\hbox{\tiny dec}}= \,a_{\hbox{\tiny nr}}$,
\,indicates the level at which dark matter interactions start modifying
significantly the damping of the primordial dark matter fluctuations.
A dark matter particle with a mass of 
$\,\sim\, 1$ MeV for example
\,(assuming appropriate dilution or annihilations to have occurred,
for a suitable relic abundance),\,
which would normally have been considered as leading to 
relatively small damping compared to the allowed scale of 100 kpc,
may actually become a warm dark matter candidate, 
as a result of the enhancement of the free-streaming scale
in region II, by a factor $\,\sqrt{\,a_{\rm dec}/a_{\rm nr}}\,$.

\vskip .2truecm

In region III the enhancement factor of the free-streaming scale,
$\,\sqrt{\,a_{\rm eq}/a_{\rm nr}}\,$,
~may be as large as $\,\sim 10^3$ in the case of 
a $\,\sim$ 10 MeV particle, which may then
acquire a just-acceptable damping length of $\,\sim 100$ kpc
\,(as compared to \linebreak
$\sim$ 100 pc for a particle of the same mass, 
in region I, according to equation (\ref{lfs3})).

\vskip .2truecm

Beyond that, we can propose a further redefinition of the notions 
of cold, warm and hot dark matter by considering
not only the free-streaming scale alone
\,(even modified by the effects of the interactions), 
\,but the whole damping scale obtained
by combining the collisional (self-damping plus induced damping) 
and free-streaming scales. In this way we 
take into account the various possible dark matter interactions, 
both with itself and with other particle 
species such as photons and neutrinos. 
CDM will then correspond to values of the 
damping scales significantly smaller than $\,l_{\hbox{\tiny struct.}}$, 
\,and WDM to values nearly as large as the allowed value.

\vskip .18truecm

Before turning to induced-damping effects, 
for which the damping of dark matter fluctuations is due to the transport 
of energy and momentum by other particle species, 
we refer to Fig. 1 of Ref.~\cite{bfs}.
\,In this plane dark matter interaction rate 
\,($\,\Gamma_{\hbox{\tiny{dm}}}\,a^3\,$)\, at decoupling 
\,versus dark matter particle mass $m_{\hbox{\tiny{dm}}}$ 
(times $\,\kappa_{\hbox{\tiny{dm}}}$),
\,we represent the regions excluded\,\footnote{One should 
also not forget the familiar relic density constraints. This normally 
excludes regions of the parameter space corresponding to heavy 
particles too weakly coupled to annihilate 
sufficiently, so that their resulting relic abundance would be too large
(unless they have been appropriately 
diluted by some prior inflation phase). \,This is particularly important 
in a large part of region I, corresponding to heavy dark matter particles 
that would decouple before getting non-relativistic 
\,-- in contrast with the standard case of heavy neutralinos for example,
which normally decouple after getting non-relativistic 
and belong to region II.}
by both self-damping and free-streaming effects.
These excluded regions may be viewed as corresponding to extensions 
of ``hot dark matter'' scenarios to the new cases 
for which {\it collisional damping} as well as 
{\it interaction-enhanced free-streaming effects} are taken into account. 
The regions relatively close 
to the excluded ones, for which the damping scale is not much lower 
than the scale $\,l_{\hbox{\tiny struct.}}\,$ of the smallest primordial 
objects, may be considered as corresponding to extensions 
of ``warm dark matter'' scenarios.

\section{Induced-damping by photons or neutrinos}
\label{sec:ind}

\vskip -.1truecm

One should also consider the additional damping effects 
induced by other particle species to which dark matter may be coupled, 
such as the photons and the neutrinos.
The magnitude of these effects is given by the 
induced-damping scale associated with a species 
$i$, as expressed by eq.\,(\ref{lid}), the integral over time running 
until the decoupling time of dark matter from this species $i$.

\vskip .2truecm
Photons and neutrinos may be particularly relevant,
not only because they are relativistic ($v_i = c$), 
\,but also because they could interact long enough 
with dark matter, 
if their interaction cross-sections with dark matter particles 
turned out to be sufficiently large. 
The corresponding induced-damping scales,
given by
\vspace{-.2truecm}
\be
\label{lid3}
l_{\hbox{\scriptsize \,photon-induced damping}}
^{\,2}
\ \ \approx \ \ \ 
\int^{t_{\hbox{\tiny \,dec.}}\hbox{\tiny\,(dm-$\gamma$)}} 
\hbox{\footnotesize $\displaystyle
\frac{\rho_\gamma}{\rho}
$}
\ \ \frac{c^2\ dt}{\Gamma_\gamma(t)\ \,a^2(t)}\ \ \ ,
\ee
for photons, and similarly for neutrinos,
should both be smaller 
than the length $\,l_{\hbox{\scriptsize struct.}}$, 
\,here taken to be $\,\sim\,$ 100 kpc.
~In particular, the constraint
$\ l_{\hbox{\tiny \,photon-ind. damp.}}$
$\simle\,100 \,$ kpc\,
leads us to upper limits on the interaction rates of dark matter
particles with photons at the corresponding decoupling time, 
and therefore to {\it \,upper limits\,}
on the dark-matter/photon interaction cross-sections.
(The more dark matter particles interact with those of a given species, 
the longer the latter can influence their properties, 
and the larger is the resulting contribution to the 
induced-damping length of the primordial dark matter fluctuations.)
The moment at which the photons cease to communicate their damping to the 
dark matter fluid should occur before the recombination epoch in order to 
avoid prohibitive damping effects.
The damping induced by neutrinos is normally expected to be very small, 
excepted in special situations for which dark matter would decouple 
from neutrinos at a temperature $\,T<1\,$ MeV. 
In that case ``freely-propagating neutrinos'' could
influence the properties of dark matter and induce a non-negligible
``collisional damping'' 
of its primordial fluctuations.
We refer the reader to Refs.\,\cite{bfs,b2}\,
for a more detailed discussion of these photon or neutrino-induced
damping effects.

\vspace{ -.05truecm}

\end{document}